\documentclass[doublecol]{epl2} 
\def\be{\begin{equation}}
\def\ee{\end{equation}}

\def\bc{\begin{center}} 
\def\ec{\end{center}}
\def\bea{\begin{eqnarray}}
\def\eea{\end{eqnarray}}

\newcommand{\Avg}[1]{\left\langle{#1}\right\rangle}
 
\title{Interdisciplinary and physics challenges of network theory}
\shorttitle{Interdisciplinary and physics challenges of network theory} %Insert here a short version of the title if it exceeds 70 characters

\author{Ginestra Bianconi}
\shortauthor{Ginestra Bianconi}

\institute{School of Mathematical Sciences, Queen Mary University of London, London E1 4NS, United Kingdom}


\abstract{Network theory has unveiled the underlying structure of complex systems such as the Internet or the biological networks in the cell. It has identified universal properties of complex networks, and the interplay between their structure and  dynamics.
After almost twenty  years of the field, new  challenges lie ahead.
These challenges concern the multilayer structure of most of the networks, the formulation of a network geometry and topology, and the development of a quantum theory of networks.
Making progress on these aspects of network theory can open new venues to address interdisciplinary and physics challenges including progress on brain dynamics, new insights into quantum technologies, and quantum gravity. 
}
\begin{document}

\maketitle

\section{Introduction}
Network theory has emerged almost twenty  years ago, as a new field for characterizing interacting complex systems, such as the Internet, the biological networks of the cell, and social networks.
It is now time to   reflect on the maturity of the field, indicating the main results obtained so far and the big challenges that lie ahead.

Initially,  the physics perspective, in particular the statistical mechanics approach, has dominated the field of Network Theory \cite{Barabasi_review, Newman_book,Doro_book,Boccaletti2006,Caldarelli_book,Havlin_book,Santo}. This point of view has played a central role to characterize the universal properties of the structure of   {\it complex networks}.  
It has been found that despite the diversity of complex networks, ranging from the Internet to the protein interaction networks in the cell,  most networks obey universal properties: they are  small-world \cite{WS}, they are scale-free \cite{BA}, and they have a non trivial community structure \cite{Santo}. 
Moreover, over the years, special attention has been addressed to the interplay between network structure and dynamics \cite{crit,Dyn}.  In fact phase diagrams of critical phenomena and dynamical processes change drastically when the dynamics is  defined on complex networks. Complex networks are responsible for significant changes in the critical behaviour of    percolation, Ising model, random walks, epidemic spreading, synchronization, and  controllability of networks \cite{crit,Dyn,Control,Control2}.

The need to characterize complex systems, to extract relevant information from them, and to understand how dynamical processes are affected by network structure, has never been more severe than in the XXI century when we are witnessing a Big Data explosion  in social sciences, information and communication technologies  and in biology. 

Under different points of view, it can be argued that network theory, started from the perspective of the theoretical physicists with the goal of answering these questions,  is becoming an increasingly  multidisciplinary field. As knowledge and amount of data about biological networks, social networks or infrastructures, are becoming more substantial,  it seems inevitable that different types of networks require different expertise, involving  scientists of the relative specific disciplines in the first place. Therefore, while network tools are becoming widely accepted in system biology, social sciences, or in engineering, the different sub-fields are becoming more specialized.

Here I am advocating that the theoretical approach is nevertheless fundamental to address the Big Challenges that lie ahead, and that physicists and mathematicians continue to be essential for the advance of the field.
We will never make real progress in the understanding of  the brain function \cite{Bullmore} or of the origin of life if we do not integrate the biological knowledge with a physics understanding \cite{Physics_brain,Goldenfeld} of these two most striking examples of emergent phenomena. In this effort, considering the brain under the point of view of multilayer networks \cite{PhysReports,Kivela,Goh_review,Makse,Plamen,Latora} could contribute to the global understanding of brain function as due to the dynamics occurring on several interacting networks. 
Moreover, to  understand the origin of life it would be crucial to  shed light on the formation of the minimal cellular networks of protocells. To this end, it would be  essential to model  how the cell has  evolved as  the most beautiful example of multilayer networks, formed by several interacting and interdependent networks. 

A very crucial question, from the theoretical physics perspective is  whether network theory is a field  that  might have a significant impact also in more traditional fields of mathematics and physics. From this point of view, two main directions are attracting the attention of an increasing number of scientists: network geometry and topology and  a quantum theory of networks. These two branches of network theory not only represent theoretical challenges, but they are likely to  have several important practical applications.
 
 Developing a new theory of network geometry and topology   could contribute to a deeper understanding of network structure, and could be crucial for solving problems in community detection and data mining \cite{Mahoney1,Vaccarino1,Vaccarino2}. Moreover,   it is believed that this theory could be  fundamental for proposing new routing strategies for packets in the Internet, solving in this way a problem of scalability of the presently used technology \cite{Kleinberg, Boguna_navigability,Boguna_Internet,Saniee}.
Additionally characterizing brain geometry and topology will contribute to a deeper understanding between brain structure, dynamics  and function \cite{brain_geometry,Ballerini,Munoz1,Munoz2,Makse2,Plamen}. Finally, geometrical network models \cite{Emergent,Quantum,Manifold,Boguna_hyperbolic,Boguna_growing} have been able to generate networks sharing the phenomenology of most complex networks and therefore provide the best way of understanding how all the universal properties of complex networks might emerge at the same time. 

A quantum theory of networks, combining  quantum mechanics and  complex networks properties,  could play a pivotal role in the  future development of quantum communication technologies \cite{quantum_Internet}. It is known that future quantum communication technologies can  improve the security and the transfer rate of current classical communication systems. When fully implemented on the large scale, it is likely that they will share some of the complexity properties of  the current communication systems.
 Therefore the cross-disciplinary field between complex network and quantum information is gaining increasing attention.
  On one side, quantum dynamical processes are increasingly  explored on network structures \cite{Mulken,Mulken2,Q0,Q1,Q2,Q3,Q4,Q5,Q6,Faccin}. On the other side, quantum information proposals \cite{Cirac,Calsamiglia},  are pushing the frontier of our understanding of how quantum networks could be realized. 
 Moreover this field  has contributed to the formulation of new entropy measures for quantifying the complexity of networks \cite{Severini,Vonneuman,Garnerone_entropy,Hancock1}, and of new ranking algorithms \cite{Garnerone_ranking,Jesus_ranking}.

Defining geometrical complex networks, and relating them  to quantum  states can open new scenario for cross-fertilization between network theory and quantum gravity.
 As Penrose wrote {\em My own view is that ultimately physical laws should find their most natural expression in terms of essentially combinatorial principles,[$\ldots$]. Thus, in accordance with such a view, should emerge some form of discrete or combinatorial spacetime} \cite{quantum_spacetime}.
At the moment, most quantum gravity approaches agree that the quantum space-time has a discrete, network-like structure \cite{Rovelli,CDT1,Dowker,Rideout}.
 Moreover, it is not to be excluded a priori that network theory, developed to understand complexity and biological systems could bring new insight on some aspects of quantum gravity.
In the words of Lee Smolin {\em A theory of quantum cosmology cannot be logically consistent if it does not describe a complex universe.}\cite{Smolin_lifeofcosmos}.

In this direction, new results have been obtained. On one side, the connection between hyperbolic complex networks and causal sets \cite{Dowker,Rideout} used in quantum gravity has been explored in a recent paper \cite{Cosmology}, and causal sets have been used to analyse complex networks \cite{Evans}.
On the other side, significant progress has been made exploring the relations between emergent network geometries, evolution of quantum network states, and quantum statistics using equilibrium \cite{graphity,graphity_Hamma} and non-equilibrium approaches \cite{Emergent,Quantum,Manifold}.
 
In the following I will focus on several topics of significant recent interest in network theory framing the results obtained so far and their possible role for solving the big interdisciplinary and physical challenges that I have here outlined.

\section{Multilayer networks}
From the cell, to the brain most networks are multilayers \cite{PhysReports,Kivela,Goh_review,Havlin1}, i.e. they are formed by several interacting networks. For example, in the cell, the protein-protein interaction network, the signaling networks, the metabolic networks and the transcription networks are not isolated but interacting, and the cell is not alive if anyone of these networks is not functional. In the brain, understanding the relation of functional and structural networks \cite{Bullmore}, forming a multilayer network,  is of fundamental importance.  Moreover, there are additional multiple ways to characterize brain networks as  multilayer structures that capture other aspects of its complexity. For example it is possible to distinguish  between the synaptic and electrical connectivity of the fully annotated brain of the worm {\em c. elegans},\cite{Latora} or it is possible to construct a multilayer network formed  of different functional network modules of the brain \cite{Makse}.

Multilayer networks have been first introduced in the context of social sciences \cite{Wasserman} to describe different types of social ties. Social network remain at the moment one of the typical examples of multilayer networks, nevertheless multilayer networks have attracted a significant interdisciplinary interest only in the last five years, because it has becoming clear that characterizing multilayer networks is fundamental to understand most complex networks including cellular networks, the brain, complex infrastructures, and economical networks in addition to social networks.
\begin{figure}
\begin{center}
\includegraphics[width=0.8\columnwidth]{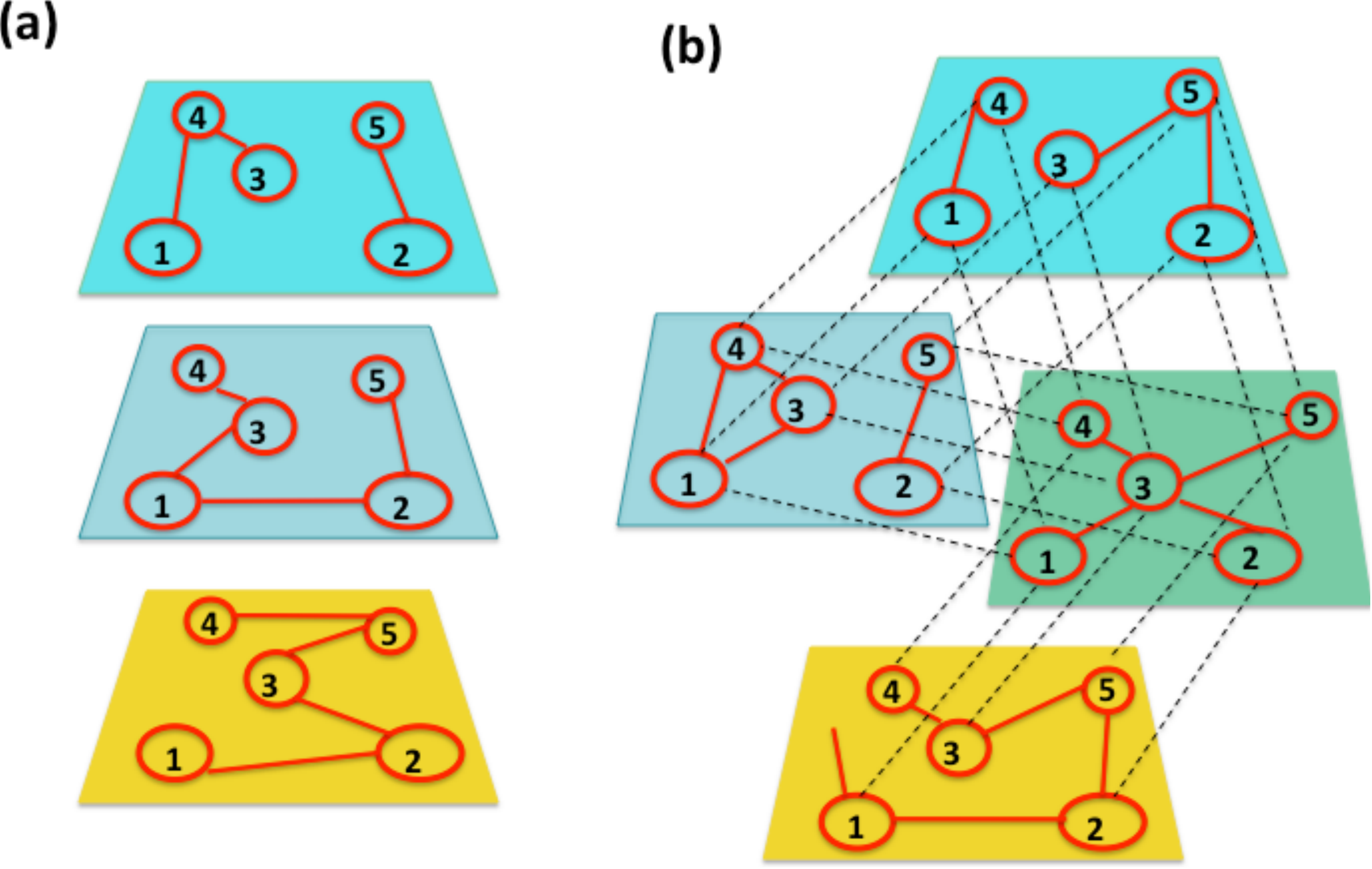}
\end{center}
\caption{(Color online) Visualization of different types of multilayer networks. Panel (a) shows a multiplex network where the same set of nodes is linked in different layers. For example these can correspond in social networks to people connected by different means of communication. Panel (b) shows an example of network of networks, where the nodes of different networks are connected by interlinks (dashed lines).}
\label{figure1}
\end{figure}

A multilayer network is not to be confused with a larger network including all the interactions. As a network ultimately is a way to encode information \cite{Anand2009} about the underlying complex system, there is a significant difference between considering all the interactions at the same level, or including the information on the different nature of the different interactions.
In a multilayer network, each interaction has a different connotation, and this property is correlated with other structural characteristics, allowing  network scientists to extract significant more information from the complex system under investigation. Therefore significant impact of this research is expected on network medicine that requires the integration of many different data regarding the patients.
 
Multilayer networks can be distinguished in two major classes: multiplex networks and network of networks (see Figure $\ref{figure1}$). A multiplex network is a network formed by the same set of nodes interacting through different type of networks (also called layers) as for example a set of people interacting through different means of communication.  Examples of  characterized multilayer networks include collaboration networks \cite{Menichetti,Latora}, transportation networks \cite{Boccaletti}, social networks \cite{Thurner}, just to name few of the most studied datasets.
Network of networks, are instead networks that are interacting with each other but are formed by different types of nodes, such as the Internet, the power-grid, and other types of interdependent infrastructures \cite{Havlin1}; or different biological networks in the cell. The links joining nodes of different layers are also called {\em interlinks}.

In these last five years the focus of the  research  has been  on multilayer network structure and dynamics \cite{PhysReports}.
It has been shown that considering the multilayer nature of  networks can modify significantly the conclusions reached by considering  single networks. A number of  dynamical processes, including percolation\cite{Havlin1,Doro_multiplex,Son,Havlin_prl,BD1,BD2,Gao1,JSTAT}, diffusion\cite{diffusion}, epidemic spreading\cite{Boguna_epidemics} and game theory \cite{game} present a phenomenology that is unexpected if one consider the layers in isolation.
Moreover, it has been shown that multilayer networks are characterized by significant correlations \cite{PRE,Latora,growing1,growing2,Goh_correlations,Mason_M} in their structure that can change the dynamical properties of the multilayer network.

\begin{figure}
\begin{center}
\includegraphics[width=0.75\columnwidth]{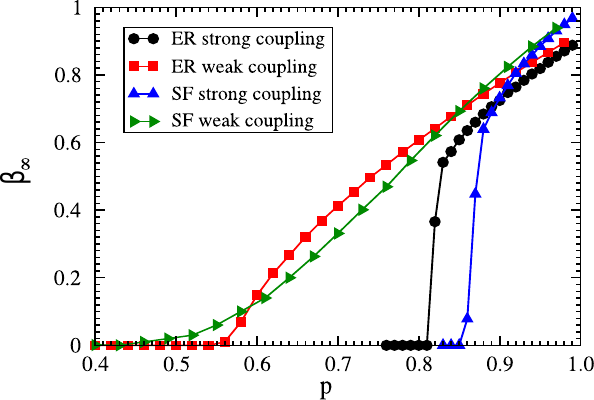}
\end{center}
\caption{(Color online) The figures show the size of the mutually connected giant component $\beta_{\infty}$, as a function of $p$, the fraction of not damaged nodes, for Erd\"os and Renyi multiplex networks (ER) and scale free multiplex networks (SF). The transition is discontinuous in the case in which most of the interlinks imply  interdependencies (strong coupling) and continuous if the fraction of interdependent interlinks is below a given threshold (weak coupling). Figure from Ref.\cite{Havlin_prl}. Copyright (2010) by The American Physical Society..}
\label{figure2}
\end{figure}

\begin{figure*}
\begin{center}
\includegraphics[width=1.85\columnwidth]{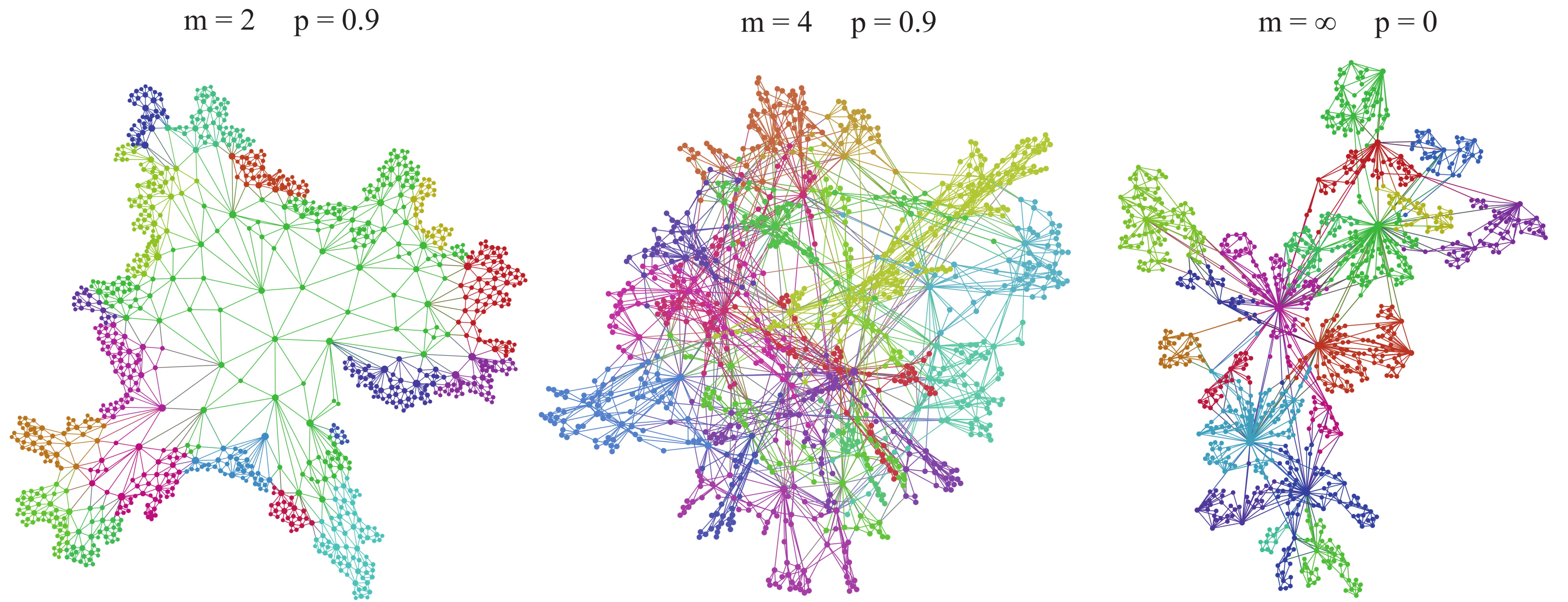}
\end{center}
\caption{(Color online) Visualization of the emergent network geometries generated by the non-equilibrium model presented in Ref.\cite{Emergent}. If at most  $m=2$ triangles are incident to a link, the model generates a 2d manifold, with small world properties and exponential degree distribution. If instead, every link can be incident to any number of triangles ($m=\infty$) the network is scale-free, small-world, has high modularity and high clustering coefficient. The intermediate case $m=4$ has broad degree distribution, high clustering,  high modularity  and is small-world.  Figure  from Ref. \cite{Emergent}.}
\label{figure3}
\end{figure*}

Particularly noticeable has been the finding that when nodes of different networks are interdependent with each other, multilayer networks might be much more fragile than single networks and may have cascading failures that yield abrupt transitions \cite{Havlin1,Havlin_prl,Son,Doro_multiplex,BD1,BD2,Gao1} . Therefore this result explains why global infrastructures are prone to dramatic avalanches of failures. In presence of interdependencies, a new type of percolation phase transition can be defined \cite{Havlin1} in which the order parameter is the size of the {\em mutually connected giant component}, i.e. an appropriate generalization of the giant component defined on single networks. When nodes of an interdependent multilayer networks are damaged with an increasing probability, the mutually connected component has a hybrid phase transition \cite{Havlin1,Doro_multiplex}, in which the size of the mutually connected giant component has a discontinuity and the system undergoes an avalanche of failures. 
If the  interdependency is only partial, i.e. some interlinks do not imply interdependencies, the mutually connected component can emerge at a continuous second order phase transition \cite{Havlin_prl} (see Figure $\ref{figure2}$).
This generalized percolation transition has been studied on multiplex networks, multilayer network and network of networks finding a rich phenomenology \cite{Havlin1,Havlin_prl,Son,Doro_multiplex, BD1,BD2,Gao1,JSTAT}. 
Multilayer networks found in biological systems \cite{Makse}, are different from man-made multilayer infrastructures, and they display a significant robustness allowing them to survive biological selection. Characterizing them could contribute to a better design of complex infrastructures.

The literature on multilayer networks is growing at a very fast rate due to the relevance of this research for a large variety of fields. Additional important results are covered in detail on the recent review articles on multilayer networks and multiplex networks\cite{PhysReports,Kivela,Goh_review}. Here we have chosen just to provide a short overview of  this topic. 

\section{Network geometry and topology}

Increasing attention has been recently addressed to the geometrical and topological characterization of networks. 
In this field, scientific research interest has been following  four major directions: characterization of the hyperbolicity of networks, formulation of emergent network geometry, characterization of brain geometry, and network topology.

{\em Characterization of the hyperbolicity of networks.}
 The characterization of the curvature of networks is a fundamental mathematical problem addressed by different mathematicians providing different alternative definitions \cite{Ollivier, Yau1,Yau2, Jost, Higuchi,Gromov}. Except from  the combinatorial curvature \cite{Higuchi} of planar graphs  there is no established consensus on the most appropriate curvature definition for network structures. Despite this fact, several approaches are used to characterize the geometry \cite{Majid}, the complexity \cite{Franzosi}, and the hyperbolicity of  networks. One way to achieve this is to measure the  Gromov $\delta$-hyperbolicity\cite{Gromov,Bary,Reka}, other ways include  embedding the network in hyperbolic spaces \cite{Aste,Boguna_navigability,Boguna_Internet},  or  in general in a Riemannian geometry \cite{Aste2,Hancock2}.

It is  believed that many complex networks have an underlying   ``hidden geometry"\cite{Boguna_Internet} and that extracting this geometry could be extremely useful.
For example, the   hidden hyperbolic geometry of the Internet  could be used to improve significantly the routing protocols, which would send the packets on a path chosen accordingly to the distance between two nodes in this  geometry \cite{Kleinberg,Boguna_navigability,Boguna_Internet}.

A noticeable series of works \cite{Boguna_hyperbolic,Boguna_growing} introduce equilibrium and non-equilibrium modelling frameworks to construct scale-free networks starting from a hidden hyperbolic geometry. In these models,  nodes are placed on the hyperbolic plane, and neighbor nodes  are more likely to be linked to each other.
The generated networks have at the same time large clustering coefficient, small-world property, scale-free degree distribution, and therefore reproduce the phenomenology observed in real network structures.
Interestingly these models can also be used to extract the hidden hyperbolic metrics of networks.

{\em Emergent network geometry.} 
The field of emergent geometry  aims at generating networks with hidden geometry without using any information of this underlying space.
This field has its origins in quantum gravity where a general problem is determining how the geometry of the continuous space-time emerges from the discrete structure that the space-time has at the Planck scale.
These models, also called pregeometric models,  where space is an emergent property of a network  or of a  
simplicial complex, have attracted large interest in quantum gravity over the years \cite{Wheeler, pregeometry2,CDT1}.
In \cite{graphity}, {\em quantum graphity} has been  proposed as an equilibrium model for emergent space-time. The model is Hamiltonian, and the low temperature phase of the network is a planar graph with some defects \cite{graphity} while the network corresponding to the high temperature regime is a complete network. 

 Recently,  manifolds and scale-free networks \cite{Emergent} with high clustering, small-world property, and a non-trivial community structure have been generated using a non-equilibrium  model of emergent geometry (see Figure $\ref{figure3}$).  This  model  is based on a growing simplicial complex formed by gluing together triangles. In this model each link can be incident at most to $m$ triangles. If $m=2$ a two dimensional manifold with exponential degree distribution is generated, if $m=\infty$ a scale-free network with high clustering, high modularity and small world distances is generated. The model is able to generate an emergent network geometry,  particularly evident in the case of the  generation of the random manifold.
Nevertheless further investigations will be needed to  specifically address the characterization of  the hidden geometry of these networks.
If instead of triangles, simplices of higher dimension, i.e. tetrahedra etc., are glued together, the growing manifolds are  scale-free for any dimension $d>2$\cite{Manifold}.

{\em Brain geometry.}
 Geometry is  fundamental to understand the brain. This is particularly true at the structural level \cite{brain_geometry}, and can have impact for developing future cortex transplants \cite{Ballerini}. Moreover, the study of  the interplay between structural and functional brain networks \cite{Bullmore}, is recently attracting large attention. It is believed that modularity \cite{Makse2,Munoz1,Munoz2} plays a role of special importance, together with the small world property, in generating a dynamical phase of frustrated synchronization, where synchronization is sustained but not stationary. 
It is possible that in the future, advances in the understanding of the geometrical  organization of the brain networks will allow to fully identify  the structural properties that favour brain dynamics.

 {\em Topology of networks.} 
 The topology of networks is attracting large attention \cite{Farber,Kahle} and the topological  characterization of network datasets and dynamical network models, is becoming a new tool of network theory. In particular a new way to define a filtration on weighted networks has been recently proposed \cite{Vaccarino1,Vaccarino2}, where the filtrations correspond to different thresholds imposed to the weights of the links. The topological analysis that results is able to extract new information from network datasets that is not possible to extract using other less recent techniques of network theory.  Finally the topological analysis of dynamical processes as epidemic spreading, can reveal the underlying topology of the network over which the spreading occurs \cite{Mason}.

\begin{figure}
\begin{center}
\includegraphics[width=.8\columnwidth]{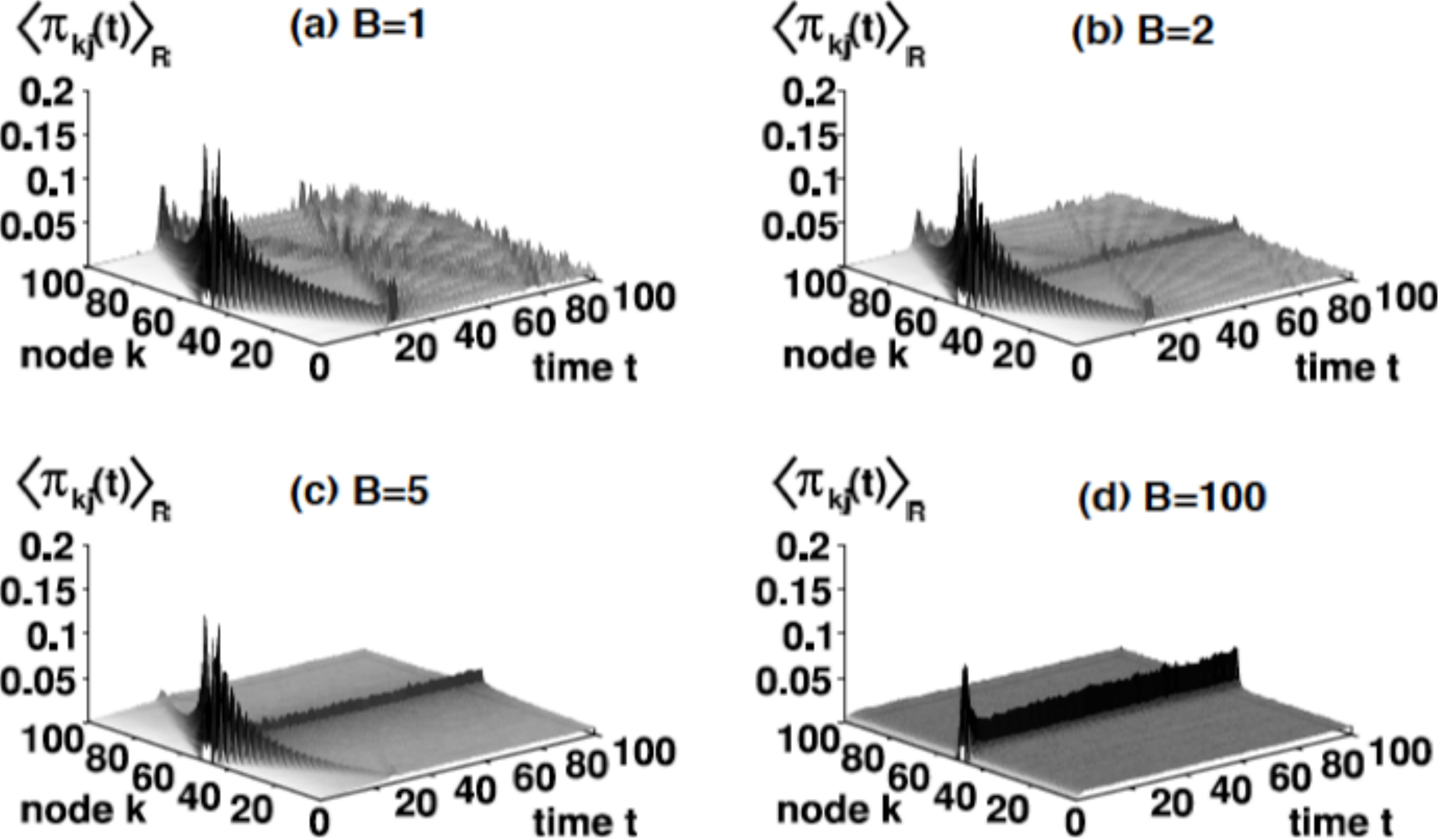}
\end{center}
\caption{The average transition probability $\Avg{\pi_{kj}(t)}$ of a Continuous-Time Quantum Random Walk on a small world network as a function of time. The initial node is node $j=50$,  the small world networks is formed by $100$ nodes placed on a ring and by  $B$ of additional random links. From Ref.\cite{Mulken2}.}
\label{figure4}
\end{figure}

\section{Quantum theory of networks}

 Characterizing the behaviour of quantum dynamics on networks is attracting increasing attention because it is relevant for new quantum technologies using  cold atoms or light-matter interaction. Extensive research has been addressed to the characterization of quantum walks on complex networks (see Figure $\ref{figure4}$) \cite{Mulken,Mulken2,Faccin} and to quantum transport \cite{Huelga}. 
Other critical phenomena studied on networks are the quantum Ising model, the Bose Hubbard model, Anderson localization, Bose-Einstein condensation between others\cite{Q0,Q1,Q2,Q3,Q4,Q5,Q6}. It has been  found  that network structure strongly affect the phase diagram of quantum dynamical processes. These results could open the venue for exploring optimal design of networks for obtaining desired dynamical properties of the quantum process under investigation. 

From the quantum information perspective, constructing quantum networks for quantum communications \cite{quantum_Internet} is certainly one of the major long-term goals. For this purpose quantum networks where the nodes are entangled, provide the ideal setup to establish quantum communication over large distances  guaranteeing more security and efficiency  than classical communication technologies. In this context, the properties of quantum random networks \cite{Cirac} have been investigated and the effect of complex networks topologies in favouring   the possibility to establish entanglement between long-distance nodes, has been characterized \cite{Calsamiglia}.

Quantum information methods have been also used to propose new quantum information entropy measures for assessing the complexity of networks\cite{Severini,Vonneuman,Hancock1,Garnerone_entropy}. In particular, the Von Neumann entropy of networks \cite{Severini} is defined by interpreting the Laplacian matrix of the networks, normalized by the total number of links, as a density matrix of a quantum state. Interestingly this entropy measure, for random scale-free networks, can be mapped to the Shannon entropy of scale-free network ensembles \cite{Vonneuman}.  In the same spirit, with the goal of characterizing classical complex networks with quantum methods,  a series of works has been devoted to the proposal of quantum algorithms for ranking nodes \cite{Garnerone_ranking,Jesus_ranking,Caldarelli_WWW}. 

Finally there is evidence for  a surprising relation between the evolution of complex growing networks and quantum statistics. This  relation was found already in the early days of the field of network science the framework of the  Bianconi-Barab\'asi model \cite{Fitness,Bose}. In fact complex networks growing according to  preferential attachment and {\em energies} of the nodes  (related to their fitness) might display a Bose-Einstein condensation, where one node grabs a finite fraction of all the links. A similar phase transition can occur also on   weighted networks \cite{Weighted}  undergoing also the condensation of the weight of the links. Growing  Cayley trees  with energies of the nodes, can be mapped to Fermi gas and follows Fermi-Dirac statistics \cite{Fermi}.  The models in \cite{Bose} and \cite{Fermi} following respectively Bose and Fermi distributions have underlying symmetries as discussed in \cite{Complex}.
Moreover,  simple or weighted networks described by equilibrium statistical mechanics follow quantum statistics \cite{Garlaschelli}. 
Recently, using models formulated for describing emergent geometry \cite{Emergent}, it was found that the relation between network and quantum statistics extend also to manifolds and to networks build starting from simplicial complexes\cite{Quantum,Manifold}. Interestingly, using similar methods used in \cite{graphity}, it has been shown that these network evolutions represent single histories of the evolution of quantum network states. These are quantum network states that can be decomposed in states associated to the nodes, and to the faces of the simplicial complex. These results deepen the understanding about the relation between complex networks and quantum statistics. In fact, the quantum network states include fermonic and bosonic occupation numbers, and their average over the networks follow respectively  Bose and Fermi statistics, also if the networks and the quantum network states do not obey equilibrium statistical mechanics.

\section{Conclusions}
Network science has had a fabulous development in the last twenty years, and is having a huge impact on a multitude of fields, from neuroscience and cell biology to economics, and social sciences.
It is now crucial for network scientists to investigate multilayer networks, characterizing the interactions between different networks. This field will have impact on a number of applications because most networks in biology, technology, economics,engineering,  or social sciences are not isolated but interacting.
Moreover, an important new challenge for network science is the full development of a network geometry and network topology, that will represent not only a big step ahead in the comprehension of discrete geometries, but will also have important practical implications for data mining, community detection and routing problems.
Finally, novel quantum technologies will require a full control of quantum dynamics on network structure, and is it likely that they will require sophisticated network design to achieve
desired properties. It is therefore of fundamental importance to fully characterize quantum dynamics on networks.
Finally, combining network geometry with quantum theory of networks could open new venues for  cross-fertilization between network theory and quantum gravity.

\end{document}